\documentclass[copyright]{eptcs}
\providecommand{\event}{DCFS 2010} 
\usepackage{breakurl}             

\usepackage{latexsym}
\usepackage[dvips]{graphics}
\usepackage{epsfig}

\def\bbbn{{\rm I\!N}} 
\def\vs{\vspace{3mm}}
\def \prend{\vrule depth-1pt height7pt width6pt}
\def \proof{\bigbreak\noindent{\bf Proof.\ \ }}
\def \endpf{{\ \ \prend \medbreak}}

\usepackage{amssymb}
\usepackage{graphicx}

\begin{document}

\newtheorem{theorem}{T\/heorem}[section]
\newtheorem{apptheo}{T\/heorem}[section]
\newtheorem{corollary}{Corollary}[section]
\newtheorem{definition}{Definition}[section]
\newtheorem{lemma}{Lemma}[section]
\newtheorem{applem}{Lemma}[section]
\newtheorem{example}{Example}[section]
    \newtheorem{appexample}{Example}[section]
\newtheorem{fact}{Fact}[section]
\newtheorem{claim}{Claim}[section]
\newtheorem{proposition}{Proposition}
\newtheorem{remark}{Remark}[section]
\newcommand{\propersubset}{\subset}
\newtheorem{open}{Open problem}

\title{Transformations Between
Different Types of Unranked Bottom-Up  Tree Automata}

\author{Xiaoxue Piao \qquad\qquad Kai Salomaa
\institute{School of Computing, Queen's University\\
Kingston, Ontario K7L 3N6, Canada}
\email{\{piao, ksalomaa\}@cs.queensu.ca}
}
\def\titlerunning{Unranked Tree Automata}
\def\authorrunning{X. Piao \& K. Salomaa}
%
%
%
%

\maketitle

\begin{abstract}
We consider the representational state complexity  of unranked
tree automata.
The bottom-up computation of
an unranked tree automaton may be either deterministic or nondeterministic,
and further variants arise depending on
whether the horizontal string languages defining the transitions are
represented by a DFA or an NFA.
Also, we consider for unranked tree
automata the alternative  syntactic definition
of determinism  introduced
by Cristau et al. (FCT'05, Lect. Notes Comput. Sci. 3623, pp. 68--79).
 We establish upper and lower bounds  for 
the state complexity of conversions
between different types of unranked tree automata.\\
Keywords: tree automata, unranked trees, state complexity,
nondeterminism
\end{abstract}

\section{Introduction}
\label{intro}

Descriptional complexity, or state complexity, of finite automata has 
been extensively studied in recent years, 
see~\cite{HK,LMSY,Sa,Yu} and references
listed there. On the other hand, very few papers explicitly discuss
state complexity of tree automata. For classical tree automaton
models operating on ranked trees~\cite{CDG,GS} many state complexity
results are  similar to corresponding results on string automata.
For example, it is well known that 
determinizing an $n$ state nondeterministic
bottom-up tree automaton gives an automaton with
at most $2^n$ states.

Modern applications of tree automata, such
as XML document processing~\cite{MSV,Sc}, use automata operating
on unranked trees. One approach is to first
encode
the unranked trees
 as binary trees~\cite{CNT}.
The other approach that we consider here
is to define the computation of the tree automaton directly on unranked
XML-trees~\cite{BMW,CDG,Sc}. 
The set of transitions of an unranked tree automaton is,
in general, infinite and the transitions are usually specified
in terms of a regular language. 
Thus, in addition to the finite set of states used
in the bottom-up computation,
 an unranked tree automaton needs
for each state $q$ and input symbol $\sigma$ a finite string automaton to
recognize the {\em horizontal language\/} consisting
of strings of states defining the transitions associated
to $q$ and $\sigma$.

Here we consider bottom-up (frontier-to-root) unranked tree automata.
Roughly speaking, we get  different models depending on whether
the bottom-up computation 
is nondeterministic or deterministic and
whether the horizontal languages are recognized 
by an NFA or a DFA ((non-)deterministic finite automaton).
Furthermore,
there is more than one way to define determinism
for unranked tree automata and we compare here
 two of the variants. 

The more common 
definition~\cite{CDG,Sc}
 requires that for any input symbol $\sigma$ and
two distinct states $q_1$, $q_2$, the horizontal languages associated,
respectively,
with $q_1$ and $\sigma$ and with $q_2$ and $\sigma$ are disjoint.
The condition guarantees that the bottom-up computation assigns
a unique state to each node. 
To distinguish this from the syntactic definition of
determinism of~\cite{CLT,RB}, we call
a deterministic tree automaton where the horizontal
languages defining the transitions
are specified by  DFAs, a {\em weakly deterministic tree automaton.}
Note that a computation of a weakly deterministic
automaton still needs to ``choose'' which
of the DFAs (associated with different states) is used to process
the sequence of states that the computation reached at the children
of the current node -- since the intersection of 
distinct horizontal languages
is empty the choice is unambiguous, however, when beginning to process
the sequence of states the automaton has no way of knowing which
DFA to use.

A different definition, that we call {\em strong determinism,}
was considered in~\cite{CLT,RB}.\footnote{
The paper \cite{CLT}
refers to weak and strong determinism, respectively, as semantic and
syntactic determinism.}
A strongly deterministic automaton associates to each input symbol a single
DFA $H_\sigma$ equipped with
 an output function, the state at a parent node
labeled by $\sigma$ is determined
(via the output function) by the state $H_\sigma$ reaches
after
processing the sequence of states corresponding to the children.
Strongly deterministic automata can be minimized efficiently and
the minimal automaton is unique~\cite{CLT,RB}. On the other hand,
interestingly it was shown in~\cite{MN}
that for weakly deterministic tree automata the minimization problem
is NP-complete and the minimal automaton need not be unique.

We study the state complexity of determinizing
different variants of nondeterministic tree automata. That is,
we develop upper and lower bounds for the size of deterministic tree
automata that are equivalent to given nondeterministic automata.
We define the size of an unranked tree automaton 
as a pair of integers
consisting of the number  of  states used in the bottom-up 
computation,
 and the sum of the sizes of the
NFAs definining the horizontal languages. Note that the
two types of states play very different roles
in computations of the tree automaton. The other possibility would
be, as is done e.g. in~\cite{MN}, to count simply the total number
of all states in the different components. 

Also, we study the state complexity of the conversions between
the strongly and the weakly deterministic tree automata. Although the former
model can be viewed to be more restricted, there exist tree languages
for which the size of a strongly deterministic automaton is smaller
than the size of the minimal weakly deterministic automaton.
It turns out to be more difficult to establish  lower bounds
for the size of weakly deterministic automata than is the case for
strongly deterministic automata. Naturally, this can be expected due
to the intractability of the minimization of weakly deterministic
automata~\cite{MN}. 

It should be noted that there are many other deterministic
automaton models used for applications on unranked trees, such
as stepwise tree automata~\cite{CNT,CGL} and nested word
automata~\cite{AM,PS}. Size comparisons between, respectively, stepwise tree
automata and strongly deterministic automata or automata operating
on binary encodings of unranked trees can be found in~\cite{MN}.
Much work remains to be done on state complexity of tree automata.

To conclude we summarize
the contents of the paper. In Section~\ref{prelim}
we recall definitions for tree  automata operating on unranked
trees and introduce some notation. In Section~\ref{D+D} we
study the descriptional complexity of conversions between
the strongly and the weakly deterministic tree automata, and
in Section~\ref{N+D} we study the size blow-up of converting
different variants of nondeterministic tree automata
to strongly and weakly deterministic
automata, respectively.
Many of the proofs have been omitted in this extended abstract for
the DCFS proceedings.

\section{Preliminaries}
\label{prelim}

We assume that the reader is familiar with the basics of
formal languages and finite automata~\cite{HU,Yu}. 
Below we briefly recall some definitions for tree automata operating
on unranked trees and fix notations. More details
on unranked tree automata and references can be found in~\cite{CDG,Sc}.
A general reference on tree automata operating on ranked trees
is~\cite{GS}.

Basic notions concerning trees, such as the root,
a leaf, a subtree, the height of a tree and children of a node
  are assumed to be
known. 
The set of non-negative
integers is $\bbbn$. 
A   {\em tree domain\/}  is a prefix-closed subset $D$ of $\bbbn^*$
such that if $ui \in D$, $u \in \bbbn^*$, $i \in \bbbn$ then
$uj \in D$ for all $j < i$. The set of nodes of a tree $t$
is represented in the well-known way as a tree domain
 ${\rm dom}(t)$ and the  node labeling 
is given by a mapping ${\rm dom}(t)
\rightarrow \Sigma$ where $\Sigma$ is a finite alphabet
of symbols.
Thus, we use labeled
ordered unranked
trees. Each node of a tree 
 has a finite number of children
with a linear order, but there is no a priori upper bound on the
number of children of a node. The set of all $\Sigma$-labeled
trees is $T_\Sigma$.

We introduce the following  notation for trees.
For $i \geq 0$, $a \in \Sigma$ and $t \in T_\Sigma$, we
denote by $a^i(t)=a(a(...a(t)...))$  a tree, where
  the  nodes $\varepsilon$, $1$, \ldots, $1^{i-1}$
 are labelled by $a$ and the subtree at node $1^i$ is $t$. 
When $a\in\Sigma$, $w=b_1b_2...b_n \in \Sigma^*$, $b_i\in\Sigma$, 
$1\leq i\leq
n$, we use $a(w)$ to denote the tree $a(b_1, b_2, ..., b_n)$. 
When $L$ is a set of strings, $a(L) = \{ a(w) \; \mid \; w \in L \}$.
The
set of all $\Sigma$-trees where exactly one leaf is labelled by a
special symbol $x$
($x \not\in \Sigma$)
is $T_\Sigma[x]$. For $t \in T_\Sigma[x]$
and $t' \in T_\Sigma$, $t(x \leftarrow t')$ denotes the
tree obtained from $t$ by replacing the unique  occurrence
of variable $x$
by $t'$. 

A {\em nondeterministic (unranked) tree automaton\/} (NTA) is a
tuple 
$A=(Q,\Sigma,\delta,F)$,
where $Q$ is the finite set of
states, $\Sigma$ is the alphabet labeling
nodes of input trees, $F \subseteq Q$ is the set of
final states, and $\delta$ is a mapping from $Q\times\Sigma$ to the
subsets of $Q^*$ which satisfies the condition that,
for each $q \in Q$, $\sigma \in \Sigma$, $\delta(q, \sigma)$ is a
regular language. 
The language $\delta(q, \sigma)$ is called
the {\em horizontal language\/} associated with $q$ and $\sigma$. 

A computation of $A$ on a tree $t \in T_\Sigma$ is a
mapping $C : {\rm dom}(t) \rightarrow Q$ such that 
 for $u \in {\rm dom}(t)$,
if $u \cdot 1, \ldots u \cdot m$, $m \geq 0$, are the
children of $u$ then 
$C(u \cdot 1) \cdots C(u \cdot m) \in \delta(C(u), t(u))$.
In case $u$ is a leaf the condition means that $m = 0$ and
$\varepsilon \in \delta(C(u), t(u))$.

Intuitively, if a computation of
$A$ has reached the children of a $\sigma$-labelled node $u$ in a
sequence of states 
$q_1,q_2, \ldots, q_m$, the computation may
nondeterministically assign a state $q$ to the node $u$ provided that
$q_1 q_2 \cdots q_m\in\delta(q, \sigma)$. 
For $t \in T_\Sigma$, $t^A \subseteq Q$ denotes the set of states
that  in some bottom-up computation $A$ may reach at the root of $t$.
The {\em tree language  recognized by $A$\/}  is
defined as $L(A) = \{ t \in T_\Sigma \mid t^A \cap F \neq \emptyset \}$.

For a tree automaton $A = (Q, \Sigma, \delta, F)$, we denote
by $H^A_{q,\sigma}$, $q \in Q$, $\sigma \in \Sigma$,
 a nondeterministic finite automaton (NFA) on strings
recognizing the horizontal language
$\delta(q, \sigma)$. 
The NFA $H^A_{q,\sigma}$ is called a horizontal automaton,
and states of different horizontal automata are called
collectively {\em
horizontal states.} 
We refer to the states of 
$Q$ that are
used in the bottom-up computation as {\em vertical states.} 

A tree automaton $A = (Q, \Sigma, \delta, F)$ is said to be (semantically)
{\em 
deterministic\/} (a DTA) if for  $\sigma \in \Sigma$ and
any two states $q_1\neq q_2$,
$\delta(q_1, \sigma) \cap \delta(q_2, \sigma)=\emptyset$.

We get a further refinement of classes of
automata depending on whether
the horizontal languages are defined using 
DFAs or NFAs.
We use NTA($M$) or DTA($M$),
respectively, to denote (the class of) nondeterministic
or deterministic  tree automata where the horizontal
languages are specified by the elements in class $M$. 
For example,
NTA(DFA) denotes the  tree automata where the horizontal
languages are recognized by a DFA. 

Note that when referring
to a tree automaton $A = (Q, \Sigma, \delta, F)$  it is always assumed that
the relation
$\delta$ is specified in terms of
automata $H^A_{q, \sigma}$, $q \in Q$, $\sigma \in \Sigma$, and 
by saying that $A$ is an NTA(DFA) we  indicate that each
$H^A_{q, \sigma}$ is a DFA.
We refer to DTA(DFA)'s also as {\em weakly deterministic tree automata\/}
to distinguish them from the below notion of strong determinism.

If $A$ is a DTA(NFA), for any tree $t \in T_\Sigma$ 
the bottom-up computation of $A$ assigns
a unique vertical state to the root of $t$, 
that is, $t^A$ is 
 a singleton
set or empty.
If the horizontal automata $H^A_{q, \sigma}$ are
DFAs, furthermore, for each transition the sequence
of horizontal states is processed deterministically. However, as discussed
in Section~\ref{intro}, a computation that has reached children
of a $\sigma$-labeled node in a sequence of states $w \in Q^*$
still needs to make the choice which of the DFAs $H^A_{q, \sigma}$,
$q \in Q$, is used to process $w$. For this reason we consider
also the following  notion
introduced in~\cite{CLT} that we call strong determinism.

A  tree automaton $A=(Q,\Sigma,\delta,F)$ is said to be
{\em strongly deterministic\/} if for each $\sigma \in \Sigma$, the
transitions are defined by a single DFA augmented with an output
function as follows.
 For $\sigma \in \Sigma$ define
\begin{equation}
\label{sdta}
H^A_{\sigma}
=(S_\sigma, Q, s_\sigma^0, F_\sigma,\gamma_\sigma,\lambda_\sigma),
\end{equation}
 where
$(S_\sigma,Q, s_\sigma^0,F_\sigma,\gamma_\sigma)$ is a DFA  with
set of states $S_\sigma$ where
$s^0_\sigma \in S_\sigma$ is the start state, $F_\sigma \subseteq
S_\sigma$ is the set of final states and
$\gamma_\sigma : S_\sigma \times Q \rightarrow S_\sigma$ is
the transition function,
 and 
$\lambda_\sigma$ is a
 function $F_\sigma \rightarrow Q$. Then we
 require that for all $q \in Q$ and $\sigma \in
\Sigma$: $\delta(q,\sigma)=\{ w \in Q^* \mid
\lambda_\sigma(\gamma_\sigma(s_\sigma^0,w))=q\}$.
  Note that the definition
guarantees that $\delta(q_1, \sigma) \cap \delta(q_2, \sigma) = \emptyset$
for any distinct $q_1, q_2 \in Q$, $\sigma \in \Sigma$.
The class of strongly deterministic tree automata is denoted
as SDTA.\footnote{Strictly
speaking, $\delta$ is superfluous in the tuple specifying an SDTA
and the original definition of~\cite{CLT} gives instead the
automata $H^A_\sigma$, $\sigma \in \Sigma$.
We use $\delta$ in order to make the notation compatible with our
other models, and to avoid having to define
 bottom-up computations of SDTAs
 separately.}


By the size of an NFA $B$, denoted ${\rm size}(B)$, we mean the
number of states of $B$.
Because the roles played by  vertical and horizontal states,
respectively, in the computations of a tree automaton are essentially
different, when measuring the size of an automaton we count the
two types of states separately.
The size of an NTA(NFA) $A = (Q, \Sigma, \delta, F)$ is defined as
$$
{\rm size}(A) = [\;|Q|; \; \sum_{q \in Q, \sigma \in \Sigma} 
{\rm size}(H^A_{q, \sigma})\;] \;\; (\in \bbbn \times \bbbn).
$$
Using notations of~(\ref{sdta}), the size of an SDTA $A$ is defined
as the pair
of integers
${\rm size}(A) = [\;|Q|; \; \sum_{\sigma \in \Sigma} |S_\sigma|\;]$.

We make the following notational convention that allows us 
to use symbols of $\Sigma$ in the definition of horizontal
languages. Unless otherwise
mentioned, we assume that a tree automaton always assigns to each
leaf symbol labeled $\sigma$ a state $\overline{\sigma}$
that is not used anywhere else in the computation. That is,
for $\sigma \in \Sigma$ and $q \in Q$, $\varepsilon \in \delta(q, \sigma)$
 only if $q = \overline{\sigma}$, 
$\delta(\overline{\sigma}, \sigma) = \{ \varepsilon \}$ and
$\delta(\overline{\tau}, \sigma) = \emptyset$ for all 
$\sigma, \tau \in \Sigma$, $\sigma \neq \tau$. 
When there is no confusion, we denote also
$\overline{\sigma}$ simply by $\sigma$.  When the alphabet $\Sigma$
is fixed, there is only a constant number of
 the special states $\overline{\sigma}$ and
 since, furthermore, the special states have the same function in
all types of tree automata, for simplicity, we do not include
them when counting the vertical states.
The purpose of this
convention is to improve readability: many 
of our constructions become
more transparent when alphabet
symbols can be used explicitly to define horizontal languages.
The convention does not  change  our state complexity
bounds that are generally given within a multiplicative constant.

To conclude this
section we give two lemmas that provide lower bound estimates for
vertical and horizontal states of SDTAs, respectively.
The lower bound condition for vertical states applies, more generally,
for DTA(NFA)'s, however, obtaining lower bounds for the number of horizontal
states  of weakly deterministic automata turns out to be more problematic.

\begin{lemma}
\label{cltv} 
Let $A$ be an SDTA or a DTA(NFA) with a set of vertical states $Q$ 
recognizing a tree language $L$.
Assume $R = \{ t_1, \ldots, t_m \} \subseteq T_\Sigma$ 
where for any $1 \leq i
<  j \leq m$ there exists  $t \in T_\Sigma[x]$ 
 such that $t(x \leftarrow t_i) \in L$ iff $t(x
\leftarrow t_j) \not\in L$. Then $|Q| \geq |R| - 1$.
\end{lemma}

\begin{lemma}
\label{clth}  
Let $A$ be an SDTA  with a set of vertical states $Q$
recognizing a tree language $L$.
Let $S$ be a finite set of tuples of $\Sigma$-trees
and let $b \in \Sigma$.
Assume that for any distinct tuples $(r_1, \ldots, r_m)$, $(s_1,
\ldots, s_n) \in S$ there exists  $t \in T_\Sigma[x]$ 
 and a sequence of trees $u_1, \ldots, u_k$
such that
\begin{equation}\label{cltheq}
t(x \leftarrow b(r_1, \ldots, r_m, u_1, \ldots, u_k)) \in L \; \mbox{ iff } \;
t(x \leftarrow b(s_1, \ldots, s_n, u_1, \ldots, u_k)) \not\in L
\end{equation}
Then the horizontal automaton $H^A_b$ needs at least $|S| - 1$ states.
\end{lemma}

\section{Size comparison of the strongly and weakly deterministic tree 
automata}
\label{D+D}

Here we give upper and lower bounds for the size of a
weakly deterministic automaton (a DTA(DFA)) simulating
a strongly deterministic one (an SDTA), and vice versa.
The computation of a DTA(DFA) can, in some sense, nondeterministically
choose which of the horizontal DFAs it uses at each transition.
An SDTA does not have this capability and it can be expected that,
in the worst case, an SDTA may need considerably more states
than an equivalent DTA(DFA). However, there exist also tree
languages for which an  SDTA can be considerably more succinct than
a DTA(DFA).

\subsection{Converting an SDTA to a DTA(DFA)}

We show that an SDTA can be quadratically smaller than
a DTA(DFA). This can be compared with~\cite{MN} where
it was shown that deterministic stepwise tree automata
can be quadratically smaller than SDTA's (that are called
 dPUTA's in \cite{MN}).

The upper bound for the conversion is expected but we include a short
proof.
In the below lemma (and afterwards)
we use  ``$\leq$'' to compare pairs of
integers componentwise. 
As introduced in Section~\ref{prelim},
 for an SDTA $A$ we
denote the deterministic
automata for the corresponsing horizontal languages  by
 $H^A_\sigma$, $\sigma \in \Sigma$.

\begin{lemma}
\label{cmu} Let $A=(Q,\Sigma,\delta,F)$ be an arbitrary SDTA.

We can construct an equivalent DTA(DFA) $A'$
where
\begin{equation}
\label{tatta1}
{\rm size}(A') \leq 
[ \; |Q|; \; 
|Q|\times \sum_{\sigma \in \Sigma} {\rm size}(H^A_\sigma) \;].
\end{equation}
\end{lemma}

\proof
For $\sigma \in \Sigma$ denote the components of $H^A_\sigma$ 
as in~(\ref{sdta}).
 Construct an equivalent
DTA(DFA) $A'=(Q,\Sigma,\delta',F)$, where for each  $\sigma \in
\Sigma$, $q\in Q$, $\delta'(q, \sigma)
= \{ w \in Q^* \mid 
\lambda_\sigma(\gamma_\sigma(s_\sigma^0,w))=q \}$.
The languages $\delta'(q_1, \sigma)$ and $\delta'(q_2, \sigma)$,
$q_1 \neq q_2$ are always disjoint, and $\delta'(q, \sigma)$ is
recognized by a DFA obtained from $H^A_\sigma$ by choosing
as the set of final states $\lambda_\sigma^{-1}(q)$, $q \in Q$,
$\sigma \in \Sigma$. The construction does not
change the number of vertical states and~(\ref{tatta1}) holds.
\endpf

Next we give a lower bound for the 
conversion. 

\begin{lemma}
\label{cmlow} Let $n,z\in \bbbn$ and choose
$\Sigma = \{ a, b, 0, 1 \}$. There exists an SDTA $B$
with input alphabet
$\Sigma$, $n$ vertical states and  $z+4n$ horizontal states,
such that any DTA(DFA) for the tree
language $L(B)$ has at least $n$ vertical states and
$n(\lfloor \log n \rfloor + 2 + z)$ horizontal states.
\end{lemma}

Using   
Lemma~\ref{cmlow} with $z = n - \lfloor \log n \rfloor $, we 
see that the upper bound of Lemma~\ref{cmu} is tight within
a multiplicative constant. This is stated as:

\begin{theorem}\label{cml}
An SDTA with $n$ vertical and $m$ horizontal states can be simulated
by a DTA(DFA) having $n$ vertical and $n \cdot m$ horizontal
states.

For $n \geq 1$, there exists a tree language $L_n$ recognized by an
SDTA with $n$ vertical and $O(n)$ horizontal states such
that any DTA(DFA) recognizing
$L_n$ has $n$ vertical and $\Omega(n^2)$ horizontal  states.
\end{theorem}

It can be viewed as expected that in the conversion of
Theorem~\ref{cml}
the number of vertical states does not change. 
However as will be discussed later, in general, for a
DTA(DFA) it may be possible to reduce the number of horizontal
states by increasing the number of vertical states.

\subsection{Converting a DTA(DFA) to an SDTA}
 
Again we give first an upper bound for the simulation.
It is known from (\cite{MN} Proposition~24)
that the
simulation does not increase the number of vertical states.

\begin{lemma}\label{mcu}
Let $B=(Q,\Sigma,\delta,F)$ be an
arbitrary DTA(DFA), where $|Q|=n$. 
Let $H^B_{q, \sigma} = (S_{q, \sigma}, Q, s^0_{q, \sigma},
F_{q, \sigma}, \gamma_{q, \sigma})$ be a DFA for the
horizontal language $\delta(q, \sigma)$, $q \in Q$, $\sigma \in \Sigma$.

We can construct an equivalent SDTA $B'$ where
$${\rm size}(B') \leq [ \; |Q|; \; \;
\sum\limits_{ \sigma \in
\Sigma}(\prod\limits_{q \in Q}(|S_{q, \sigma}|-
|F_{q, \sigma}|)+\sum\limits_{q \in Q}|F_{q, \sigma}| \cdot
\prod\limits_{p \in Q, p \neq q}
(|S_{p, \sigma}| - |F_{p, \sigma}|)) \;]$$.
\end{lemma}

If $B$ has $m$ horizontal states, Lemma~\ref{mcu} gives for the
number of horizontal states of $B'$ a worst-case upper bound
that is less than $2^m$ but is not polynomial in $m$.
Next we give a lower bound construction.

\begin{lemma}\label{mclow}
Let $\Sigma=\{a,b,0,1\}$.
For any $m \in \bbbn$ and
relatively prime numbers $2 \leq  k_1<k_2<...<k_m$, there exists a
tree language $L$ over $\Sigma$ recognized by a DTA(DFA) $B$ with
${\rm size}(B) = [ \; m; \;\; \sum_{i=1}^m k_i + O(m\log m)
\; ]$
 such that any SDTA
recognizing $L$ has  at least $m$ vertical
states and $\Pi_{i=1}^m k_i$ horizontal states.
\end{lemma}

\proof
 Let $y_i \in \{ 0, 1 \}^*$ be the 
binary representation 
of $i \geq 1$.
We define $L= \bigcup_{1 \leq i \leq m} a^i({(b^{k_i})}^* y_i)$. 

We define for $L$ a DTA(DFA)  $B = (Q,\Sigma,\delta,F)$, where
$Q=\{q_1,...,q_m\}$, $F=\{q_1\}$, 
$\delta(a,q_i)={(b^{k_i})}^*\cdot y_i + q_{i+1}$, for $1\leq
i \leq m-1$, and $\delta(a,q_m)={(b^{k_m})}^*\cdot y_m$.
Note that the bottom-up computation of $B$ is deterministic because
different horizontal languages are marked by distinct binary
strings $y_i$.
%

Each horizontal language 
${(b^{k_i})}^*\cdot y_i + q_{i+1}$ can be recognized
by a DFA with
 $k_i
+ \lfloor \log i \rfloor +3$ states,
and in total  $B$
has
$ \sum_{i=1}^m k_i + \sum_{i=1}^m(\lfloor \log i
\rfloor) + 3m$ horizontal states (and $m$ vertical states).

Let $B'=(Q',\Sigma,\delta',F')$ be an arbitrary
SDTA recognizing $L$. By choosing $R = \{ a(b^{k_i} y_i) \mid
1 \leq i \leq m  \} \cup \{ a(b) \}$, Lemma~\ref{cltv}
gives  $|Q'| \geq m$.

We show  that the DFA $H^{B'}_a$, with notations
as in~(\ref{sdta}), defining
transitions corresponding to symbol $a$ needs at least
$\prod_{i=1}^m k_i$ states. 
Suppose 
that  $H^{B'}_a$ has less than
$\prod_{i=1}^m k_i$ states. Then there exist $0\leq j <
s < \prod_{i=1}^m k_i$ such that $H^{B'}_a$ reaches  the same state 
after reading strings $b^j$ and $b^s$, respectively. 
There must exist $1\leq r \leq m$ such that $k_r$ does not divide
$s-j$. Let $z=j+(k_r - j \; {\rm mod} \; k_r)$. 
Since $H^{B'}_a$ reaches the same state on $b^j$ and
$b^s$, it follows that $H^{B'}_a$ reaches
the same state also on $b^z \cdot y_r$ and $b^{z + s - j}\cdot
y_r$, respectively. This means that $a^{k_r}( b^z y_r )$ is
accepted by $B'$ if and only if $a^{k_r}( b^{z + s - j}\cdot
y_r )$ is accepted by $B'$, which is a contradiction because
$k_r$ divides $z$ and does not divide $z + s - j$.
\endpf

In the above proof, using a more detailed analysis
it could  be shown that $H^{B'}_a$ needs $\Omega(m \cdot \log m)$
 additional states
to process the strings $y_i$, however, this would not change
the worst-case lower bound.



Now we establish that the upper and lower bounds for the
DTA(DFA)-to-SDTA conversion are
within a multiplicative constant, at least when the sizes of the
horizontal DFAs are large compared to the number of vertical states.

\begin{theorem}
\label{oatta1}
An arbitrary DTA(DFA) $B = (Q, \Sigma, \delta, F)$ has an
equivalent SDTA $B'$ with 
\begin{equation}
\label{eq1}
{\rm size}(B') \leq
[ \; |Q|; \;
\sum_{\sigma \in \Sigma} \prod_{q \in Q} {\rm size}(H^B_{q, \sigma}) \;],
\end{equation}
 and,
for an arbitrary
 $m \geq 1$ there exists a DTA(DFA) $B = (Q, \Sigma, \delta, F)$
with $|Q| = m$ such that for any equivalent SDTA $B'$ 
the size of $B'$ has a lower bound
 within a multiplicative constant of~(\ref{eq1}).
\end{theorem}

\proof
The upper bound follows from
Lemma~\ref{mcu}.
We get the lower bound from Lemma~\ref{mclow} by choosing
 each $k_i$ to be at least $m \cdot \log m$, $i = 1, \ldots, m$.
\endpf

We note that when converting a DTA(DFA) $B = (Q, \Sigma, \delta, F)$ to
an equivalent SDTA $A$, for each $\sigma \in \Sigma$ the
horizontal DFA $H^{A}_\sigma$ needs at least as many states
as a DFA recognizing $L_{B, \sigma} = \bigcup_{q \in Q} \delta(q, \sigma)$.
Note that from $H^{A}_\sigma$ we obtain a DFA for $L_{B, \sigma}$ simply
by ignoring the output function. 
However, $H^{A}_\sigma$ needs to provide more detailed information
for a given input string than a DFA simply recognizing $L_{B, \sigma}$,
and in fact $H^{A}_\sigma$ recognizes the marked union, as
formalized below, of
the languages $\delta(q, \sigma)$.

We say that a DFA $A = (Q, \Sigma, s_0, F, \gamma)$ 
equipped with an output
function $\lambda: F \rightarrow \{ 1, \ldots, m \}$
 recognizes
the {\em marked union\/} of pairwise
disjoint regular languages $L_1$, \ldots,
$L_m$, if $L_i = \{ w \in \Sigma^* \mid \lambda(\gamma(s_0, w)) = i \}$,
$i = 1, \ldots, m$.
The following result establishes that
 the state complexity of marked union may be arbitrarily much
larger than the state complexity of union.
\begin{proposition}
\label{cltunion}
Let $A = (Q, \Sigma, s_0, F, \gamma, \lambda)$ be a DFA with
output function $\lambda: F \rightarrow \{ 1, \ldots m \}$
that recognizes the marked union of disjoint
languages $L_i$, $i = 1, \ldots, m$,
and let $B$ be the minimal DFA for $\bigcup_{i=1}^m L_i$. 

Then 
 ${\rm size}(A) \geq {\rm size}(B)$,
and for any $m \geq 1$
there exist disjoint regular languages $L_i$, $1 \leq i \leq m$,
 such that
${\rm size}(B) = 1$ and the   ${\rm size}(A) \geq m$.
\end{proposition}

\section{Converting nondeterministic tree automata to deterministic
automata}
\label{N+D}

In this section we consider 
conversions of different variants of nondeterministic
automata into equivalent strongly and weakly deterministic
automata. 

\subsection{Converting a nondeterministic automaton to an SDTA}


\begin{lemma}\label{nncu}
Let $A = (Q, \Sigma, \delta, F)$ be an NTA(NFA)
and for $q \in Q$, $\sigma \in \Sigma$ denote ${\rm size}(H^A_{q,
\sigma}) = m_{q, \sigma}$.

\begin{description}
\item[{\rm (i)}]
We can construct an equivalent SDTA $B$  where
\begin{equation}
\label{eq2}
{\rm size}(B) \leq
[ \; 2^{|Q|}; \;
\sum\limits_{\sigma \in \Sigma} 2^{(\sum\limits_{q\in Q} m_{q, \sigma})}
\; ].
\end{equation}

\item[{\rm (ii)}]
If $A$ is a DTA(NFA), in the upper bound~(\ref{eq2}) the number
of vertical states is at most $|Q|$.

\end{description}
\end{lemma}
We do not require the automaton to be complete and, naturally, in~(\ref{eq2})
the number of vertical states of $B$ could be reduced to $2^{|Q|} - 1$. 
A similar small
improvement could be made to the number of horizontal states,
but it would
make the formula look rather complicated.

Also,
in Lemma~\ref{nncu}~(ii) the upper bound for the number of horizontal
states could be slightly reduced using a more detailed analysis,
as in the proof of Lemma~\ref{mcu}, that takes into account
that, in no situation, two distinct  NFAs defining the
horizontal languages associated with a fixed input symbol $\sigma$
can accept simultaneously. 

Lemma~\ref{nncu} did not discuss the case where the bottom-up
computation is nondeterministic but the horizontal languages
are represented in terms of DFAs.
We note that for an NTA(DFA) $A = (Q, \Sigma, \delta, F)$ 
the construction used in the proof
of Lemma~\ref{nncu} gives for the size of an equivalent
SDTA only the  upper bound~(\ref{eq2}).
Although the horizontal languages of $A$ are defined using
DFAs, the horizontal languages of the equivalent SDTA $B$ are over
the alphabet ${\cal P}(Q)$, and this means that the upper bound
for the number of horizontal states would not be improved.

Next we state two lower bound results.

\begin{lemma}
\label{ndclow}
Let $\Sigma = \{ a, b \}$.
For any relatively prime numbers $m_1,m_2,...,m_n$, there exists a
tree language $L$ over $\Sigma$
such that $L$ is recognized by an NTA(DFA) $A$ with
${\rm size}(A) \leq [ \; n; \;
(\sum\limits_{i=1}\limits^n m_i) + 2n- 2 \;],$
 and any SDTA for $L$ needs at least $2^n - 1$ vertical states
and $(\prod\limits_{i=1}\limits^n m_i) - 1$ horizontal states.
\end{lemma}

\begin{lemma}\label{dncgen}
For $n\geq 1$, there exists a tree language $L_n$ 
recognized by a DTA(NFA) $A$ with
$n$ vertical and less than $n\log n$ horizontal states
such that for any
SDTA $B$ for $L_n$, ${\rm size}(B)
\geq [ \; n; \; 2^n \; ]$. 
\end{lemma}

The lower bounds given by the above two lemmas are far away from the
corresponding upper bounds in Lemma~\ref{nncu}. Furthermore, 
we do not have a worst-case construction for 
general NTA(NFA)'s that would provably  give  an essentially better
lower bound than the one obtained for NTA(DFA)'s in Lemma~\ref{ndclow}.

\subsection{Converting a nondeterministic automaton to a DTA(DFA)}

We begin with a simulation result establishing an upper
bound.

\begin{lemma}
\label{oatta2}
Let $A = (Q, \Sigma, \delta, F)$ be an NTA(NFA)
and for $q \in Q$, $\sigma \in \Sigma$ denote ${\rm size}(H^A_{q,
\sigma}) = m_{q, \sigma}$.

\begin{description}
\item[{\rm (i)}]
There exists a  DTA(DFA) $B$  equivalent to $A$ where
\begin{equation}
\label{guasian}
{\rm size}(B) \leq
[ \; 2^{|Q|}; \;  2^{|Q|} \cdot
(\sum_{\sigma \in \Sigma} 
 2^{(\sum_{q\in Q} m_{q, \sigma})})
\; ].
\end{equation}

\item[{\rm (ii)}]
If $A$ is a DTA(NFA), it has an equivalent DTA(DFA) $B$ where
$$
{\rm size}(B) \leq [ \; |Q|; \;
\sum_{q \in Q}\sum_{\sigma \in \Sigma} 2^{m_{q, \sigma}} \; ].
$$

\end{description}
\end{lemma}

Roughly speaking, the simulation uses a 
standard subset construction~\cite{Yu} for the
set of vertical states, and in order to guarantee
that the bottom-up computation remains deterministic
the DFA for the horizontal language
corresponding to $P \subseteq Q$, $\sigma \in \Sigma$, needs to
simulate each horizontal NFA  of $A$ corresponding  
to $\sigma$. 
In the case where $A$ is an NTA(DFA) we do not have a significantly
better bound than~(\ref{guasian}), because the horizontal languages
of the DTA(DFA) consist of strings of subsets of $Q$, which means
that we again have to simulate multiple computations of each horizontal
DFA of $A$.
In the below lower bound
construction of Theorem~\ref{oatta4} we, in fact, use an
NTA(DFA).

We do not have a  lower bound that would match the bound
of Lemma~\ref{oatta2}. Recall that strongly deterministic automata
can be minimized efficiently and the minimal automaton is unique~\cite{CLT},
however, minimal DTA(DFA)'s are, in general, not unique and minimization
is intractable~\cite{MN}. When trying to establish lower bounds
for the size of a DTA(DFA) 
$A = (Q, \Sigma, \delta, F)$ there is the difficulty that by adding
more vertical states, and hence more horizontal languages, 
it may still be possible that the total number 
of horizontal states is reduced. For example, suppose that $A$
has a horizontal language
$\delta(q, \sigma) = (a+b)^* b(a+b)^7$,
 where the minimal DFA has 256 states.\footnote{Note
that $\delta(q, \sigma)$
is a typical example of a language where the NFA-to-DFA size blow-up
is large.} This language
can be represented as a disjoint union of 8 regular languages where the
sum of the sizes of the minimal DFAs is only 176. 
Thus, by replacing
the state $q$ by 8 distinct vertical states (that could be equivalent
in  the bottom-up computation) we could reduce the size of
$A$.

In fact, we do not have a general lower bound condition,
analogous to Lemma~\ref{clth},   for the number of horizontal
states of DTA(DFA)'s 
and the below lower bound result
relies on an ad hoc proof.

Let $\Sigma = \{ a, b \}$. Let $p_1$, \ldots, $p_n$ be the first
$n$ primes. Define the tree language
\begin{eqnarray}
\label{matta5}
T_n & = & \{ a^i (b^k) \mid  i \geq 1, \, k \geq 0, \; 
 (\exists 1 \leq
j \leq n) [ k \equiv 0 \; ({\rm mod} \; p_j ) \mbox{ and } i
\equiv j \; ({\rm mod} \; n )] \}. \nonumber
\end{eqnarray}

\begin{theorem}
\label{oatta4}
The tree language $T_n$ can be recognized by an NTA(DFA) $A$ with
${\rm size}(A) = [ \; n; \; (\sum_{i=1}^n p_i) + 2n \;]$,
and for any DTA(DFA) $B$ recognizing $T_n$,
$$
{\rm size}(B) \geq [\; 2^n - 1; \; (2^n-1) \cdot \prod_{i = 1}^n p_i \; ].
$$
\end{theorem}

\noindent
 Theorem~\ref{oatta4} gives a construction where
converting an NTA(DFA) to a DTA(DFA) causes an exponential blow-up
in the number of vertical states, and additionally the size
of each of the (exponentially many) horizontal DFAs is considerably
larger than the original DFA. However, the size blow-up of the horizontal
DFAs does not match the upper bound of Lemma~\ref{oatta2}.
In the proof of Theorem~\ref{oatta4},
roughly speaking, we  use a particular type of
unary horizontal languages
in order to be able to (provably) establish that there cannot be a trade-off
between the numbers of vertical and horizontal states, 
and with this type of constructions it seems difficult to approach
the worst-case
size blow-up of Lemma~\ref{oatta2}.

\section{Conclusion}

We have studied the  state complexity of
conversions between different models of  tree automata
operating on unranked trees.
For the conversion of  weakly deterministic automata
into  strongly deterministic
automata, and vice versa,
 we established lower bounds that are within a multiplicative constant
of the corresponding upper bound. However, for the size blow-up
of converting nondeterministic automata to
(strongly and weakly) deterministic automata the upper and lower bounds remain
far apart, and this is a topic for further research.

Since a minimal weakly deterministic automaton need not be unique~\cite{MN},
it is, in general,  hard to establish lower bounds  for 
the number of horizontal states of weakly
deterministic automata and we do not have 
tools like Lemma~\ref{clth} that is used for 
strongly deterministic automata. 
Weakly deterministic automata  can have  trade-offs
between the numbers of vertical and horizontal states, respectively, and
it would be useful  to  establish some upper bounds 
for  how much the
number of horizontal states can be reduced by introducing additional
vertical states.





\end{document}